\documentclass[12pt]{article}
\usepackage{graphicx}

\def\hybrid{\topmargin 0pt      \oddsidemargin 0pt
        \headheight 0pt \headsep 0pt
       \voffset-1cm
        \textwidth 6.25in       
       \textheight 9.5in       
        \marginparwidth 0.0in
        \parskip 5pt plus 1pt   \jot = 1.5ex}
\catcode`\@=11
\def\marginnote#1{}

\newcount\hour
\newcount\minute
\newtoks\amorpm
\hour=\time\divide\hour by60
\minute=\time{\multiply\hour by60 \global\advance\minute by-\hour}
\edef\standardtime{{\ifnum\hour<12 \global\amorpm={am}%
        \else\global\amorpm={pm}\advance\hour by-12 \fi
        \ifnum\hour=0 \hour=12 \fi
        \number\hour:\ifnum\minute<10 0\fi\number\minute\the\amorpm}}
\edef\militarytime{\number\hour:\ifnum\minute<10 0\fi\number\minute}

\def\draftlabel#1{{\@bsphack\if@filesw {\let\thepage\relax
   \xdef\@gtempa{\write\@auxout{\string
      \newlabel{#1}{{\@currentlabel}{\thepage}}}}}\@gtempa
   \if@nobreak \ifvmode\nobreak\fi\fi\fi\@esphack}
        \gdef\@eqnlabel{#1}}
\def\@eqnlabel{}
\def\@vacuum{}
\def\draftmarginnote#1{\marginpar{\raggedright\scriptsize\tt#1}}

\def\draftlabel#1{{\@bsphack\if@filesw {\let\thepage\relax
   \xdef\@gtempa{\write\@auxout{\string
      \newlabel{#1}{{\@currentlabel}{\thepage}}}}}\@gtempa
   \if@nobreak \ifvmode\nobreak\fi\fi\fi\@esphack}
        \gdef\@eqnlabel{#1}}
\def\@eqnlabel{}
\def\@vacuum{}
\def\draftmarginnote#1{\marginpar{\raggedright\scriptsize\tt#1}}

\def\draft{\oddsidemargin -.5truein
        \def\@oddfoot{\sl preliminary draft \hfil
        \rm\thepage\hfil\sl\today\quad\militarytime}
        \let\@evenfoot\@oddfoot \overfullrule 3pt
        \let\label=\draftlabel
        \let\marginnote=\draftmarginnote
   \def\@eqnnum{(\theequation)\rlap{\kern\marginparsep\tt\@eqnlabel}%
\global\let\@eqnlabel\@vacuum}  }

\def\numberbysection{\@addtoreset{equation}{section}
        \def\theequation{\thesection.\arabic{equation}}}

\def\underline#1{\relax\ifmmode\@@underline#1\else
        $\@@underline{\hbox{#1}}$\relax\fi}

\def\titlepage{\@restonecolfalse\if@twocolumn\@restonecoltrue\onecolumn
     \else \newpage \fi \thispagestyle{empty}\c@page\z@
        \def\thefootnote{\fnsymbol{footnote}} }

\def\endtitlepage{\if@restonecol\twocolumn \else  \fi
        \def\thefootnote{\arabic{footnote}}
        \setcounter{footnote}{0}}  
\relax

\numberbysection
\hybrid

\newfont{\Bbb}{msbm10 scaled 1\@ptsize00}
\newfont{\Bbbb}{msbm7 scaled 1\@ptsize00}

\newcommand{\DDD}{\raise-1pt\hbox{$\mbox{\Bbbb D}$}}

\newcommand{\UUU}{\raise-1pt\hbox{$\mbox{\Bbbb U}$}}

\newcommand{\ZZ}{\mbox{\Bbb Z}}
\newcommand{\z}{\raise-1pt\hbox{$\mbox{\Bbbb Z}$}}

\newcommand{\sss}{\raise-1pt\hbox{$\mbox{\Bbbb S}$}}

\def\beq{\begin{equation}}
\def\eeq{\end{equation}}
\def\p{\partial}

\newtheorem{lemma-definition}{Lemma-Definition}[section]

\begin{document}

\begin{titlepage}

\title{Elliptic solutions of the Toda lattice with 
constraint of type B and deformed Ruijsenaars-Schneider system}

\author{V. Prokofev\thanks{
Skolkovo Institute of Science and Technology, 143026, Moscow, Russia,
e-mail: vadprokofev@gmail.com}
\and
A.~Zabrodin\thanks{
Skolkovo Institute of Science and Technology, 143026, Moscow, Russia and
National Research University Higher School of Economics,
20 Myasnitskaya Ulitsa,
Moscow 101000, Russia and
NRC ``Kurchatov institute'', Moscow, Russia;
e-mail: zabrodin@itep.ru}}

\date{February 2023}
\maketitle

\vspace{-7cm} \centerline{ \hfill ITEP-TH-07/23}\vspace{7cm}

\begin{abstract}

We study elliptic solutions of the recently introduced Toda lattice
with the constraint of type B and derive equations 
of motion for their poles. The dynamics of poles is given by the
deformed Ruijsenaars-Schneider system. We find its commutation
representation in the form of the Manakov triple and study properties
of the spectral curve. By studying more general
elliptic solutions (elliptic families), we also 
suggest an extension of the deformed
Ruijsenaars-Schneider system to a field theory.

\end{abstract}

\end{titlepage}

\vspace{5mm}

%

\tableofcontents

\vspace{5mm}

\section{Introduction}

The investigation of dynamics of poles of 
singular solutions to nonlinear integrable
equations was initiated in the seminal paper \cite{AMM77}, 
where elliptic and rational 
solutions to the Korteweg-de Vries and Boussinesq equations 
were studied. These studies were continued in 
\cite{Krichever78,CC77,Krichever80}, where it was shown that 
poles of elliptic solutions to the Kadomtsev-Petviashvili (KP) 
equation move as particles of the integrable Calogero-Moser (CM) 
many-body system
\cite{Calogero71}-\cite{OP81}.

The method suggested by Krichever \cite{Krichever80} 
for elliptic solutions 
of the KP equation consists in substituting the solution 
not in the KP equation itself but
in the auxiliary linear problem for it (this implies a 
suitable pole ansatz for the wave
function). This method allows one to obtain the equations 
of motion together with their
Lax representation. Later, the power of this method was demonstrated
in other important examples. In particular, 
dynamics of poles of elliptic solutions to the 2D Toda 
lattice \cite{UT84}
was studied by this method in \cite{KZ95}, see also \cite{Z19}. 
It was proved that the poles move as particles of the 
integrable Ruijsenaars-Schneider (RS) many-body system \cite{RS86,Ruij87} 
which is a relativistic generalization
of the CM system. Another important example is related to 
the B-version of the KP equation (BKP) 
\cite{DJKM81}-\cite{Tu07}. Elliptic solutions to the BKP equation 
were studied by Krichever's method in \cite{RZ20}, where equations 
of motion for the poles were obtained, together with their commutation
representation in the form of the Manakov's triple \cite{Manakov}. 

Recently, a new integrable hierarchy of the Toda type was suggested
\cite{KZ22}. It is the Toda hierarchy with the constraint of type B.
The first member of the hierarchy is the following system of equations
for two unknown functions $v$, $f$ depending on a space variable
$x$ and two time variables $t_1$, $t_2$:
\beq\label{int3}
\left \{\begin{array}{l}
\displaystyle{ \p_{t_1}\log \Bigl (v(x)v(x+\eta )\Bigr )=
\frac{f(x+\eta )}{v(x+\eta )}}-\frac{f(x)}{v(x)},
\\ \\
\p_{t_2}v(x)-\p_{t_1}f(x)=2v^2(x)\Bigl (v(x-\eta )-v(x+\eta )\Bigr ).
\end{array} \right.
\eeq
Here $\eta$ is a parameter having the meaning of the lattice spacing 
in the $x$-direction.
Let us note that a similar hierarchy was suggested earlier 
in the paper \cite{GK10} as an integrable discretization of the 
Novikov-Veselov equation but the close connection with the Toda
lattice was not mentioned there. 

Solutions to the Toda lattice with the constraint of type B can be 
expressed in terms of the tau-function $\tau (x)$
as follows:
\beq\label{int4}
v(x)=\frac{\tau (x+\eta )\tau (x-\eta )}{\tau^2 (x)},
\quad
f(x)=v(x)\, \p_{t_1}\! \log \frac{\tau (x+\eta )}{\tau (x-\eta )}.
\eeq
As is shown in \cite{KZ22},
the tau-function $\tau (x)$ of the Toda lattice with the constraint
of type B
is related to the tau-function $\tau^{\rm Toda} (x)$ of the 
2D Toda lattice as
\beq\label{int4a}
\tau^{\rm Toda} (x)=\tau (x) \tau (x-\eta ).
\eeq
This relation should be compared with the relation between tau-functions
of the KP and BKP hierarchies: the former is the square of the latter.
In the Toda case, it is not the square but the arguments of the two 
factors are shifted by $\eta$; in the limit $\eta \to 0$ 
they become the same.

An interesting problem is to find dynamics of poles of 
elliptic solutions to the Toda 
equation of type B. As in the Toda case, one considers solutions 
which are elliptic functions of $x$, with poles $x_i$ depending 
on the time $t=t_1$. This problem was recently 
addressed in \cite{KZ22a}, where
equations of motion for the poles were obtained from the 
condition that the first auxiliary linear problem for the 
equations (\ref{int3}) has meromorphic solutions. The corresponding 
many-body system turns out to be a deformation of the RS system. 
The equations of motion are:
\beq\label{i4}
\ddot x_i +\sum_{j\neq i}^N \dot x_i \dot x_j
\Bigl (\zeta (x_{ij}+\eta )+
\zeta (x_{ij}-\eta )-2\zeta (x_{ij})\Bigr )+g(U_i^--U_i^+)=0,
\eeq
where dot means the time derivative, 
\beq\label{i5}
U_i^{\pm}=\prod_{j\neq i}
\frac{\sigma (x_{ij}\pm 2\eta )
\sigma (x_{ij}\mp \eta )}{\sigma (x_{ij}\pm \eta )\sigma (x_{ij})},
\quad x_{ij}=x_i-x_j
\eeq
and $\sigma (x)$, $\zeta (x)$ are the Weierstrass $\sigma$-
and $\zeta$-functions. We use them 
throughout the paper. Their 
definitions and properties are given in Appendix A.
It is evident that the deformation parameter 
$g$ can be eliminated from the 
formulas by re-scaling the time variable as $t\to g^{-1/2}t$
(at $g=0$ (\ref{i4}) becomes the RS system). In what
follows we fix $g$ to be $g=\sigma (2\eta )$.
With this choice of $g$, equations (\ref{i4}) are exactly the 
dynamical equations for poles of elliptic
solutions to the Toda lattice with the constraint
of type B. 
The deformed RS 
system was further studied in \cite{Z22}, 
where the complete set of integrals of motion was found.

However, the method of \cite{KZ22a} does not allow one to find
any commutation representation of equations of motion and this 
remained to be an unsolved problem. This is the problem that we 
address in the present paper. Using an appropriate pole ansatz
for the $\Psi$-function (the solution to the auxiliary linear problem),
we apply the method of \cite{Krichever80} to the 
Toda lattice with the constraint of type B. In this way, we obtain
the commutation representation of equations (\ref{i4}) in the form
of Manakov's triple. We also study properties of the spectral curve
and show that the $\Psi$-function is the Baker-Akhiezer function
on the spectral curve. We show that the spectral curve admits a
holomorphic involution with two fixed points, as it should be for
algebraic-geometrical solutions to the Toda hierarchy with the 
constraint of type B.

It is known that integrable models of the CM and RS type admit 
extensions to field theories (``field analogues'') in which the
coordinates of particles $x_i$ become ``fields'' $x_i(x,t)$ 
depending not only
on the time $t$ but also on a space variable $x$. Equations of motion
of these more general models can be obtained as equations
for poles of more general elliptic solutions (called 
elliptic families in \cite{AKV02}) to nonlinear integrable equations.
In this more general case the solutions 
are elliptic functions of a 
linear combination $\displaystyle{u=\sum_k \beta_k t_k}$ 
of higher times $t_k$ of the hierarchy, their poles $u_i(x,t)$
being functions of the space and time variables $x,t$ (for example,
in the KP/CM case $x=t_1$, $t=t_2$). The equations of motion, together
with their commutation representation, can be obtained by an appropriate
pole ansatz. The
equations of motion of the field
analogues of the CM or RS systems
were obtained by this method in \cite{AKV02} and \cite{ZZ21}
respectively (see also \cite{Z22a}, where 
elliptic families of solutions to the constrained Toda lattice with
constraint of type C \cite{KZ21a} were
discussed). In this paper we apply this method to the Toda hierarchy
with the constraint of type B and obtain the field generalization 
of the deformed RS system. Note that the equations of motion were 
derived in \cite{Z23} by a different method, which does not allow one 
to find a commutation representation for them. Here we reproduce that
result and provide the commutation representation which is a
field extension
of the Manakov triple.

The paper is organized as follows. In section 2 we introduce the
pole ansatz for the solution of the auxiliary linear problem and 
obtain the equations of motion together with their commutation
representation. Properties of the spectral curve are studied in 
Section 3. Section 4 is devoted to analytic properties of the 
$\Psi$-function on the spectral curve, which characterize it as 
the Baker-Akhiezer function. In Section 5 we consider elliptic families 
of solutions to the Toda hierarchy with constraint of type B and
obtain equations of motion for the field analogue of the deformed RS
system together with their commutation representation.
Section 6 contains concluding remarks.
There are also two appendices. In Appendix A the definitions and 
properties of the special functions used in the main text are presented.
Appendix B is devoted to the proof of an important matrix identity
for elliptic Cauchy matrices.

\section{The pole ansatz}

The Toda lattice with constraint of type B follows from compatibility
conditions for auxiliary linear problems. 
The first of them is
the differential-difference equation \cite{KZ22}
\beq\label{a1}
\p_{t}\psi (x) =v(x) \Bigl (\psi (x+\eta )-\psi (x-\eta )\Bigr ),
\eeq
where $t=t_1$ and 
$v(x)$ is expressed through the tau-function $\tau (x)$ as
in (\ref{int4}).
For our purposes it is convenient to pass to another gauge and 
rewrite equation (\ref{a1}) in terms of the function 
\beq\label{a1b}
\Psi (x)=\frac{\tau (x+\eta )}{\tau (x)}\, \psi (x+\eta ).
\eeq
The equation for $\Psi$ reads:
\beq\label{a1c}
\p_t \Psi (x-\eta )=\Psi (x) +b(x)\Psi (x-\eta ) - 
u^{-}(x)\Psi (x-2\eta ),
\eeq
where
\beq\label{a1d}
b(x)=\p_t \log \frac{\tau (x)}{\tau (x-\eta )},
\quad
u^-(x)=\frac{\tau (x-2\eta )\tau (x+\eta )}{\tau (x-\eta )\tau (x)}.
\eeq

For elliptic solutions of the Toda lattice of type B, we have:
\beq\label{a2}
\tau (x)=C\prod_{j=1}^N \sigma (x-x_j).
\eeq
The zeros $x_j$ are assumed to be all distinct. It follows that
the coefficient functions
$$
b(x)=\sum_j\dot x_j 
\Bigl (\zeta (x-x_j -\eta )-\zeta (x-x_j)\Bigr ),
$$
$$
u^-(x)=\prod_j \frac{\sigma (x-x_j-2\eta )\sigma (x-x_j +\eta )}{\sigma
(x-x_j-\eta )\sigma (x-x_j)}
$$
in the equation (\ref{a1c})
are elliptic (double-periodic with periods $2\omega$,
$2\omega '$). 
Therefore, 
one can find double-Bloch solutions $\Psi (x)$, i.e., solutions such that 
$\Psi (x+2\omega )=b \Psi (x)$, $\Psi (x+2\omega ' )=b' \Psi (x)$
with some Bloch multipliers $b, b'$. Any non-trivial 
double-Bloch
function (i.e., not just an exponential function) must have poles.
The simplest non-trivial double-Bloch function having one pole 
in the fundamental domain is the function
\beq\label{Phi1}
\Phi (x, \lambda )=
\frac{\sigma (x+\lambda )}{\sigma (\lambda )\sigma (x)}\,
e^{-\zeta (\lambda )x}.
\eeq
It depends on the spectral parameter $\lambda$.
We call it the Krichever function since he 
introduced it for analysis of elliptic solutions as early as in 1980
\cite{Krichever80}. The main properties of the Krichever function are
listed in Appendix A.
We use the following pole ansatz for $\Psi$:
\beq\label{a3}
\Psi (x)=k^{x/\eta} \sum_{i=1}^N c_i \Phi (x-x_i, \lambda ),
\eeq
where the coefficients $c_i$ do not depend on $x$. The parameters
$k, \lambda$ are spectral parameters which are going to be connected by 
equation of the spectral curve. 
Using the quasiperiodicity properties (\ref{quasi}) of the Krichever
function,
we see that $\Psi$ given by (\ref{a3}) 
is indeed a double-Bloch function with Bloch multipliers
$$b=k^{2\omega /\eta}
e^{2(\zeta (\omega )\lambda - \zeta (\lambda )\omega )}, \qquad
b '=k^{2\omega ' /\eta}
e^{2(\zeta (\omega ' )\lambda - \zeta (\lambda )\omega ' )}.$$
In what follows we often suppress 
the second argument of $\Phi$ writing simply 
$\Phi (x)=\Phi (x, \lambda )$. 
The substitution of the pole ansatz into (\ref{a1c}), yields:
\beq\label{a3a}
\begin{array}{l}
\displaystyle{
k^{-1}\sum_i \dot c_i \Phi (x\! -\! x_i\! -\! \eta )-k^{-1}
\sum_i c_i \dot x_i \Phi '(x\! -\! x_i\! -\! \eta )}
\\ \\
\displaystyle{\phantom{aaaaa}=\, 
\sum_i c_i \Phi (x-x_i)+
k^{-1} \sum_j\dot x_j 
\Bigl (\zeta (x\! -\! x_j \! -\! \eta )-\zeta (x\! -\! x_j)\Bigr )
\sum_i c_i \Phi (x\! -\! x_i\! -\! \eta )}
\\ \\
\displaystyle{\phantom{aaaaaaaaaaaaaaa}
-k^{-2}
\prod_j \frac{\sigma (x-x_j-2\eta )\sigma (x-x_j +\eta )}{\sigma
(x-x_j-\eta )\sigma (x-x_j)}\sum_i
c_i \Phi (x\! -\! x_i\! -\! 2\eta ),}
\end{array}
\eeq
where dot means the $t$-derivative and 
$\Phi '(x):=\p_x \Phi (x, \lambda )$. 
Both sides have poles at the points
$x=x_i$ and $x=x_i+\eta$ 
(possible poles at $x=x_i+2\eta$ in the last terms 
cancel by zeros of the 
numerator). The second order poles at $x=x_i+\eta$ cancel identically.
Identification of the first order poles at $x=x_i$
gives the equations
\beq\label{a4}
c_i -k^{-1}\dot x_i \sum_j c_j \Phi (x_{ij}-\eta )-k^{-2}\sigma (2\eta )
U_i^- \sum_j c_j \Phi (x_{ij}-2\eta )=0,
\eeq
where
\beq\label{a4a}
U_i^-=\prod_{j\neq i}\frac{\sigma (x_{ij}-2\eta )
\sigma (x_{ij}+\eta )}{\sigma (x_{ij}-\eta )\sigma (x_{ij})}.
\eeq
Below we will also encounter the function $U_i^+$ which differs from
$U_i^-$ by the change $\eta \to -\eta$.
Introducing the $N\! \times \! N$ 
matrix $L=L(k, \lambda )$ with matrix elements
\beq\label{a6}
L_{ij}(k, \lambda )=\dot x_i \Phi (x_{ij}-\eta , \lambda )+
k^{-1}\sigma (2\eta )
U_i^- \Phi (x_{ij}-2\eta , \lambda )
\eeq
and the vector ${\bf c}=(c_1, \ldots , c_N)^{T}$,
we can write (\ref{a4}) as 
\beq\label{a6a}
L(k, \lambda ){\bf c}=k{\bf c}
\eeq
which implies that
\beq\label{a7}
\det \Bigl (kI-L(k, \lambda \Bigr )=0
\eeq
(here and below $I$ is the unity matrix).
This is the equation of the spectral curve. A point of the curve
is a pair $P=(k, \lambda )$ with $k$, $\lambda$ 
satisfying equation (\ref{a7}). Properties of the spectral curve
will be discussed below.
Identification of the first order poles at 
$x=x_i+\eta$ in (\ref{a3a}) gives
the equations
\beq\label{a8}
\dot c_i =\sum_j M_{ij}c_j \quad \mbox{or} 
\quad \dot {\bf c}=M{\bf c},
\eeq
where the matrix $M=M(k, \lambda )$ is given by
\beq\label{a9}
\begin{array}{l}
\displaystyle{
M_{ij}(k, \lambda )=\dot x_i (1-\delta_{ij})
\Phi (x_{ij}, \lambda ) +
k^{-1} \sigma (2\eta )U_i^+\Phi (x_{ij}-\eta , \lambda )}
\\ \\
\phantom{aaaaaaaaaaaaa}-
\displaystyle{
\delta_{ij}\Bigl (\sum_k \dot x_k \zeta (x_{ik}+\eta )-\sum_{k\neq i}
\zeta (x_{ik})\Bigr )}.
\end{array}
\eeq
The system of equations (\ref{a6a}), (\ref{a8}) is overdetermined.
The compatibility condition is
\beq\label{a10}
\Bigl (\dot L +[L, M]\Bigr ){\bf c}=0.
\eeq

In order to deal with the compatibility condition it is convenient
to introduce matrices
\beq\label{a11}
A_{ij}^{0}=(1-\delta_{ij})
\Phi (x_{ij}), \quad 
A_{ij}=\Phi (x_{ij}-\eta ), \quad B_{ij}=\Phi (x_{ij}-2\eta )
\eeq
and diagonal matrices
\beq\label{a12}
\begin{array}{l}
\dot X_{ij}=\delta_{ij}\dot x_i, 
\quad
U^{\pm }_{ij}=\delta_{ij}U_i^{\pm},
\\ \\
\displaystyle{
Z_{ij}^{\pm}=\delta_{ij}\Bigl (\sum_k \dot x_k \zeta (x_{ik}\pm \eta )-
\sum_{k\neq i}\dot x_k \zeta (x_{ik})\Bigr )},
\\ \\
\displaystyle{
D_{ij}^{-}=\delta_{ij}\sum_{k\neq i} \Bigl (\zeta (x_{ik}-2 \eta )+
\zeta (x_{ik}+\eta )-\zeta (x_{ik}-\eta )-\zeta (x_{ik})\Bigr )},
\\ \\
\displaystyle{
S_{ij}^{-}=\delta_{ij}\sum_{k\neq i} \dot x_k 
\Bigl (\zeta (x_{ik}-2 \eta )+
\zeta (x_{ik}+\eta )-\zeta (x_{ik}-\eta )-\zeta (x_{ik})\Bigr ).}
\end{array}
\eeq
We will also use the matrices
$
A_{ij}'=\Phi '(x_{ij}-\eta )$, $B_{ij}'=\Phi '(x_{ij}-2\eta ).
$
In this notation, the matrices $L$ and $M$ read
\beq\label{a13}
\begin{array}{l}
L=\dot X A +gk^{-1}U^-B,
\\ \\
M=\dot X A^0 +gk^{-1}U^+A -Z^+,
\end{array}
\eeq
where $g=\sigma (2\eta )$.

The calculation of the left hand side of (\ref{a10}) yields:
\beq\label{a14}
\begin{array}{l}
\dot L +[L, M]= \Bigl (\ddot X +
(Z^+ \! +\! Z^- )\dot X +g(U^- \! - \! U^+) \Bigr )A + 
gk^{-1}(U^-\!  -\! U^+)A(L -kI)
\\ \\
\phantom{aaaaaaaaaaaaaaaaaaaaaa}
+\, W_0 +gk^{-1} W_1 +g^2 k^{-2}W_2,
\end{array}
\eeq
where
\beq\label{a15}
\begin{array}{l}
W_0=\dot X^2 A' -\dot X A' \dot X +\dot XA \dot X A^{0} -
\dot X A^{0}\dot X A -\dot X A Z^+ -\dot X Z^- A,
\\ \\
\begin{array}{l}
\!\! 
W_1=U^- \dot X D B -U^- (S^- \! - \! Z^+) B +U^- \dot X B' -U^- B' \dot X
+\dot X A U^+ A 
\\ \\
\phantom{aaaaaaaaaaaaa}
+ U^- B \dot X A^0 -\dot X A^0 U^- B -
U^- B Z^+ -U^- A \dot X A,
\end{array}
\\ \\
W_2=U^- B U^+ A -U^+ A U^- B -(U^- - U^+) A U^- B.
\end{array}
\eeq
A direct calculation using the identities (\ref{Phi2}), (\ref{Phi3}),
shows that $W_0=0$. The calculation of $W_2$, with the help of 
(\ref{Phi3}), gives:
\beq\label{a16}
\begin{array}{l}
\Bigl ( U^- B U^+ A -U^+ A U^- B \Bigr )_{ij}
\\ \\
\begin{array}{l}
\displaystyle{
=\, \Phi (x_{ij}-3\eta ) \sum_k \Bigl [
U_i^- U_k^+ \Bigl (\zeta (x_{ik}-2\eta )+\zeta (x_{kj}-\eta )
-\zeta (x_{ij}-3\eta +\lambda )+\zeta (\lambda ) \Bigr )}
\\ \\
\phantom{aaaaaaaaa}
-U_i^+ U_k^- \Bigl (\zeta (x_{ik}-\eta )+\zeta (x_{kj}-2\eta )
-\zeta (x_{ij}-3\eta +\lambda )+\zeta (\lambda ) \Bigr )\Bigr ].
\end{array}
\end{array}
\eeq
Using the fact that the sum of residues of the elliptic function
$$
\Bigl (\zeta (x\! -\! x_j\! -\! \eta )-
\zeta (x\! -\! x_i \! +\! 2\eta )\Bigr )\prod _k
\frac{\sigma (x-x_k +2\eta )\sigma (x-x_k -\eta )}{\sigma (x-x_k+\eta )
\sigma (x-x_k)}
$$
is equal to zero, we see that
$$
\begin{array}{l} 
\phantom{aaaaaaa}\Bigl ( U^- B U^+ A -U^+ A U^- B \Bigr )_{ij}
\\ \\
\displaystyle{= \, 
\Phi (x_{ij}\! -\! 3\eta )(U_i^-\! -\!  U_i^+)
\sum_k \Bigl [
U_k^- \Bigl (\zeta (x_{ik}\! -\! \eta )+\zeta (x_{kj}\! -\! 2\eta )
-\zeta (x_{ij}\! -\! 3\eta \! +\! \lambda )+\zeta (\lambda ) \Bigr ),}
\end{array}
$$
so $W_2=0$. The calculation which shows that $W_1=0$ is most involved.
In this calculation, one should use the identity which follows
from the fact that the sum of residues of the elliptic function
$$
\Bigl (\zeta (x\! -\! x_j\! -\! \eta )-
\zeta (x\! -\! x_i \! +\! \eta )\Bigr )\prod _k
\frac{\sigma (x-x_k +2\eta )\sigma (x-x_k -\eta )}{\sigma (x-x_k+\eta )
\sigma (x-x_k)}
$$
is equal to zero (this function has simple poles at $x=x_k$
and $x=x_k-\eta$ and a second order pole at 
$x=x_i-\eta$).

As a result, we have the matrix identity
\beq\label{a17}
\dot L +[L, M]=R(L-kI)+P,
\eeq
where $P$ is the matrix
$$
P=\Bigl (\ddot X +
(Z^+ \! +\! Z^- )\dot X +g(U^- \! - \! U^+) \Bigr )A,
$$
and
$$
R=g(U^- \! - \! U^+) A.
$$
The compatibility condition (\ref{a10}) states that $P=0$, i.e,
\beq\label{a18}
\ddot X +
(Z^+ \! +\! Z^- )\dot X +g(U^- \! - \! U^+)=0,
\eeq
which are equations of motion (\ref{i4}) of the deformed RS system.

We have obtained a commutation representation of the equations
of motion in the form of the Manakov triple:
\beq\label{a19}
\dot L +[L, M]=R(L-kI), \quad R_{ij}=gk^{-1}(U^-_i \! - \! U^+_i) 
\Phi (x_{ij}-\eta , \lambda ).
\eeq
Introducing ${\cal L}(k, \lambda )=L(k, \lambda )-kI$ and
$$
M^{\pm}=\dot X A^0 +gk^{-1}U^{\pm}A -Z^+,
$$
we can write it in the form
\beq\label{a20}
\dot {\cal L}(k, \lambda )=M^-{\cal L}(k, \lambda )-
{\cal L}(k, \lambda )M^+,
\eeq
so the matrices ${\cal L}, M^+, M^-$ form the Manakov triple.

Note that the matrix $R$ in (\ref{a19}) is traceless: 
$$\mbox{tr}\, R =gk^{-1}\Phi (-\eta , \lambda )\sum_i (U^-_i -U^+_i)=0$$
since $\displaystyle{\sum_i (U^-_i -U^+_i)}$ is 
proportional to the sum of residues
of the elliptic function
$$
F(x)=\prod_j \frac{\sigma (x-x_j-2\eta )
\sigma (x-x_j+\eta )}{\sigma (x-x_j-\eta )\sigma (x-x_j)}.
$$
The fact that the matrix $R$ is traceless is used in the next section
for the proof that the spectral curve is an integral of motion.

\section{The spectral curve}

The equation of the spectral curve is
\beq\label{s1}
\det {\cal L}(k,\lambda )=\det \Bigl (kI -L(k, \lambda )\Bigr )=0.
\eeq
The time evolution
$L\to L(t)$ of our ``Lax matrix'' 
is not isospectral. Nevertheless, the characteristic polynomial, 
$\det (k I-L )$,
is an integral of motion, so the spectral curve does not 
depend on time. Indeed, 
$$
\frac{d}{dt}\, \log \det (L-k I)=
\frac{d}{dt}\, \mbox{tr}\log (L-k I)
$$
$$
=\, \mbox{tr}\Bigl ( \dot L(L-k I)^{-1}\Bigr )=
\mbox{tr}R =0,
$$
where we have used (\ref{a19}) and the fact 
that the matrix $R$ is traceless. The characteristic polynomial 
$\det (kI-L(k, \lambda ))$ is a generating function for integrals
of motion (strictly speaking, it is not polynomial but Laurent 
polynomial in $k$).

To study properties of the spectral curve, it is convenient to
pass to the gauge-transformed Lax matrix $\tilde L=
e^{-\eta \zeta (\lambda )}H^{-1}LH$, where
$H$ is the diagonal matrix with matrix elements 
$H_{ii}=e^{-\zeta (\lambda )x_i}$, and to the spectral parameter
$z=ke^{-\eta \zeta (\lambda )}$. Then the equation of the spectral
curve is
\beq\label{s2}
\det \Bigl (zI -\tilde L (z, \lambda )\Bigr )=0,
\eeq
where
\beq\label{s3}
\tilde L_{ij}(z, \lambda )=
\dot x_i \phi(x_{ij}\! -\! \eta , \lambda )-
gz^{-1} U_i^- \phi(x_{ij}\! -\! 2\eta , \lambda ), 
\quad 
\phi (x, \lambda ):=\frac{\sigma (x+\lambda )}{\sigma (\lambda )
\sigma (x)}.
\eeq
Let us denote
\beq\label{s2a}
Q(z, \lambda )=\frac{\det \Bigl (zI -
\tilde L (z, \lambda )\Bigr )}{\sigma (2N\eta -\lambda )}.
\eeq
It is the generating function of integrals of motion.
The equation of the spectral curve is 
\beq\label{s2b}
Q(z, \lambda )=0.
\eeq
The calculation of the determinant in (\ref{s2a}) \cite{Z22} shows that
$Q(z, \lambda )$ is given by
\beq\label{g10}
\begin{array}{c}
\displaystyle{
Q(z, \lambda )=\frac{z^N}{\sigma (2N\eta -\lambda )} -
\frac{z^{-N}}{\sigma (\lambda )}}
\\ \\
\displaystyle{\phantom{a}+\, 
\sum_{k=1}^N z^{N-k} \frac{\sigma (\lambda -k\eta )}{\sigma (\lambda )
\sigma (2N\eta \! -\! \lambda )
\sigma (k\eta )}\, J_k-
\sum_{k=1}^{N-1} z^{k-N} 
\frac{\sigma (\lambda -2N\eta +k\eta )}{\sigma (\lambda )
\sigma (2N\eta \! -\! \lambda )
\sigma (k\eta )}\, J_{k}},
\end{array}
\eeq
where $J_k$ are integrals of motion of the deformed RS system. 
They are given by \cite{Z22}
\beq\label{im7a}
J_n=\sum_{m=0}^{[n/2]} J_{n, m},
\eeq
where
\beq\label{im7}
J_{n, m}=\frac{\sigma (n\eta )}{\sigma ^{n-2m}(\eta )}
\sum_{{\cal I}, {\cal I}', {\cal I}\cap {\cal I}'=\emptyset
\atop |{\cal I}|=m, |{\cal I}'|=n-2m} \,
\Bigl (\prod_{j\in {\cal I}'}\dot x_j\Bigr )
\Bigl (\prod_{i,j\in {\cal I}' \atop i<j}
V(x_{ij})\Bigr ) \Bigl (\prod_{i\in {\cal I}}\, \prod_{\ell \in {\cal N}
\setminus ({\cal I}\cup {\cal I}')}\! W^{-}(x_{i\ell} )\Bigr ).
\eeq
Here the
summation is over 
subsets ${\cal I}$ and
${\cal I}'$ of the set ${\cal N}=\{1, \ldots , N\}$ 
such that ${\cal I}\cap {\cal I}'=\emptyset$ and
\beq\label{im7b}
V(x)=\frac{\sigma^2(x)}{\sigma (x+\eta )\sigma (x-\eta )},
\eeq
\beq\label{im7c}
W^-(x)=\frac{\sigma (x-2\eta )\sigma (x+\eta )}{\sigma (x-\eta )
\sigma (x)}.
\eeq
In particular,
\beq\label{g10a}
J_1=\sum_{i=1}^N \dot x_i.
\eeq

The characteristic equation
$Q(z, \lambda )=0$
defines a Riemann surface $\tilde \Gamma$ which is a $2N$-sheet covering
of the $\lambda$-plane. 
Any point of it is the pair
$(z, \lambda )$, where $z, \lambda$ are connected 
by equation (\ref{s2b}). There are $2N$
points above each point $\lambda$. 
It is easy to see from the right hand side of
(\ref{g10})
that the Riemann surface $\tilde \Gamma$ 
is invariant under the simultaneous transformations
\beq\label{trans}
\lambda \mapsto \lambda +2\omega, \quad z\mapsto 
e^{-2\zeta (\omega )\eta}z \quad \mbox{and} \quad
\lambda \mapsto \lambda +2\omega ', \quad z\mapsto 
e^{-2\zeta (\omega ')\eta}z.
\eeq
The factor of $\tilde \Gamma$ over the transformations (\ref{trans}) is an
algebraic curve $\Gamma$ which covers the elliptic curve ${\cal E}$ 
with 
periods $2\omega , 2\omega '$.
It is the spectral curve of the deformed RS model.

It is clear from (\ref{g10}) that 
the spectral curve $\Gamma$ admits a holomorphic involution 
$\iota$ with two fixed points. Indeed,
the equation $Q(z, \lambda )=0$ is invariant under the
involutive transformation
\beq\label{g12}
\iota : (z, \lambda )\mapsto (z^{-1}, 2N\eta -\lambda ),
\eeq
as is easily seen from (\ref{g10}). Therefore, the fixed points can lie 
above the points $\lambda_{*}$ such that
$\lambda_{*} =2N\eta -\lambda_{*}$ modulo the lattice with 
periods $2\omega , 2\omega '$, i.e.
$\lambda_{*}=N\eta -\omega _{\alpha}$, where 
$\omega_{\alpha}$ is either $0$ or one of the three
half-periods $\omega_1=\omega$, $\omega_2 =\omega '$, 
$\omega_3=\omega +\omega '$. 
Substituting this into the equation of the spectral curve, 
we conclude that
the fixed points are $Q_1=(1, N\eta )$, $Q_2=(-1, N\eta )$
and there are no fixed
points above $\lambda_{*}=N\eta -\omega _{\alpha}$ 
with $\omega_{\alpha}\neq 0$.

The genus $h$ 
of the spectral curve $\Gamma$ can be found using 
the following argument. Let us apply the Riemann-Hurwitz formula
to the covering $\Gamma \to {\cal E}$. We have $2h-2=\nu$, where $\nu$ is the number 
of ramification points of the covering, which are zeros on $\Gamma$
of the function $\p Q(z, \lambda )/\p z$. 
Since the number of zeros of a function 
on $\Gamma$ is equal to the number of its poles (counted 
with multiplicities), we can find $\nu$ as the number of poles 
of $\p Q/\p z$. It is clear that poles of this function may lie
only above the points $\lambda =0$ and 
$\lambda^{\iota}=0$, where we denote 
$\lambda^{\iota}=2N\eta -\lambda$.
As $\lambda \to 0$, we see from (\ref{g10}) that 
one root of the Laurent polynomial $Q(z, \lambda )$ tends to $\infty$
while other  $2N-1$ roots remain finite and non-zero. Similarly, as
$\lambda \to 2N\eta$, one root of the Laurent polynomial 
$Q(z, \lambda )$ tends to $0$ while other $2N-1$ roots remain non-zero.
Therefore, we can write
\beq\label{g13b}
Q(z, \lambda )=\frac{1}{\sigma (2N\eta )}\,
(z-h_N(\lambda )\lambda^{-1})
\displaystyle{\prod_{j=1}^{N-1}
(z-h_j(\lambda ))\prod_{k=1}^N (1-\tilde h_k(\lambda )z^{-1})}, 
\quad \lambda \to 0
\eeq
and
\beq\label{g13c}
Q(z, \lambda )=- \frac{1}{\sigma (2N\eta )}\, 
(z^{-1}\! -\! h_N(\lambda^{\iota})
(\lambda^{\iota})^{-1})
\displaystyle{\prod_{j=1}^{N-1}
(z^{-1}\! -\! h_j(\lambda^{\iota} ))
\prod_{k=1}^N (1\! -\! \tilde h_k(\lambda^{\iota} )z)}, \!
\quad \lambda^{\iota} \to 0,
\eeq
where the functions $h_j(\lambda )$ are regular at $\lambda =0$ and 
$h_N(0)=J_1$. From (\ref{g13b}) and (\ref{g13c}) we see that 
one sheet above
$\lambda =0$ (the one where $z=\infty$) 
and one sheet above $\lambda =2N\eta$ (where $z=0$) are distinguished.
The points $P_{\infty}=(\infty , 0)$ and $P_0=(0, 2N\eta )$ are
marked points of the spectral curve.
Taking the $z$-derivative of (\ref{g13b}), we see that the function
$\p Q/ \p z$ has a pole of order $N-1$ on the distinguished sheet
above $\lambda =0$ and simple poles on the other $N-1$ sheets.
From (\ref{g13c}) we see that the function
$\p Q/ \p z$ has a pole of order $N+1$ on the distinguished sheet
above $\lambda =2N\eta$ and simple poles on the other $N-1$ sheets.
So the total number of poles is $\nu =4N-2$, hence $h=2N$.

\section{The $\Psi$-function as the Baker-Akhiezer function on the 
spectral curve}

The coefficients
$c_i$ in the pole ansatz for the function $\Psi$ (\ref{a3})
are functions on the spectral curve $\Gamma$:
$c_i=c_i(t, P)$ ($P=(z, \lambda )$ is a point on the curve).
Let us normalize them by the condition $c_1(0, P)=1$.
After normalization the components
$c_i(0,P)$ become meromorphic functions on $\Gamma$ outside the 
marked points
$P_{\infty}$ and $P_0$ located
above $\lambda =0$ and $\lambda =2N\eta$. 
The location of poles of the $c_i$'s depends on the initial data.
As we shall see, the function $\Psi$ has essential singularities 
at the marked points.

Let ${\cal S}(t)$ be 
the fundamental matrix of solutions to the equation 
$\p_t {\cal S} =M{\cal S}$, ${\cal S}(0)=I$. It is a regular function of 
$z, \lambda$ for $\lambda \neq 0, \, 2N\eta$.
Using the Manakov's triple representation (\ref{a19}), we can write
$$
\Bigl (\dot L +[L,M]+R(L-k I)\Bigr ){\bf c}(t)=0.
$$
Using the relations ${\bf c}(t)={\cal S}(t){\bf c}(0)$ and $M=\dot {\cal S} {\cal S}^{-1}$, 
we rewrite this equation
as
$$
\Bigl [\p_t \Bigl ({\cal S}^{-1}(L-k I){\cal S}\Bigr ) +
R(L-kI){\cal S}\Bigr ]{\bf c}(0)=0.
$$
Equivalently, we can represent it in the form of the differential equation
$$
\p_t {\bf b}(t)={\cal W}(t){\bf b}(t), 
\qquad {\cal W}(t)={\cal S}^{-1}R{\cal S},
$$
for the vector
$
{\bf b}(t)={\cal S}^{-1}(L-k I){\bf c}(t)$ 
with the initial condition ${\bf b}(0)=0$. Under mild conditions which 
are satisfied in our case in the general position, 
the differential equation with zero initial data
has the unique solution
${\bf b}(t)=0$ for all $t>0$.
It then follows that ${\bf c}(t)={\cal S}(t){\bf c}(0)$ is the 
common solution of the equations $\dot {\bf c}=M{\bf c}$ 
and $L{\bf c}=k{\bf c}$
for all $t>0$.
Therefore, the vector ${\bf c}(t, P)$ has the same $t$-independent poles as 
${\bf c}(0,P)$.

The number of these poles can be found using the argument that the
$N$-particle deformed RS system is equivalent to a reduction of the
$2N$-particle RS system in which the particles stick together in pairs
with the distance $\eta$ between the particles in each pair. As is known
from \cite{KZ95}, the number of poles of the $\Psi$-function for the
latter system is $2N-1$. The reduction means that the spectral curve 
becomes special (it admits the involution) but the number of poles 
of the $\Psi$-function remains the same.

In order to investigate the analytic properties 
of the $\Psi$-function, it is convenient to 
pass to the gauge equivalent pair of matrices
$\tilde L$, $\tilde M$, where
$$
\tilde L=e^{-\eta \zeta (\lambda )}H^{-1}LH , 
\quad \tilde M(z) = -H^{-1}\p_t H +H^{-1}M(k)H
$$
with the diagonal matrix $H_{ij}=\delta_{ij}e^{-\zeta (\lambda )x_i}$ 
as before. 
The gauge-transformed linear system is
$$
\tilde L \tilde {\bf c}=z\tilde {\bf c}, \qquad
\p_t \tilde {\bf c}=\tilde M \tilde {\bf c}, 
$$
where 
$\tilde {\bf c}=H^{-1}{\bf c}$.  

Let us first consider what happens above the point $\lambda =0$.
Near this point we can expand:
\beq\label{psi1}
\tilde L_{ij}(z, \lambda )=\lambda^{-1}\dot x_i +gz^{-1}\lambda^{-1}
U_i^{-} +\ell_{ij} +O(\lambda ), \quad \lambda \to 0,
\eeq
where $\ell_{ij}$ is some matrix. Therefore, for solutions of the 
eigenvalue equation $\tilde L \tilde {\bf c}=z\tilde {\bf c}$
we should consider separately two cases. One is 
$\displaystyle{\sum_j \tilde c_j \neq 0}$ at $\lambda =0$, then 
$$
\tilde c_i = \dot x_i +O(\lambda ), \quad z=\lambda^{-1}\sum_j \dot x_j
+O(1)
$$
(here we use another normalization than $\tilde c_1 =1$). 
This case corresponds to the distinguished sheet of the spectral curve
(near the point $P_{\infty}$). The other is
$\displaystyle{\sum_j \tilde c_j =\lambda +O(\lambda^2)}$ as
$\lambda \to 0$, then we obtain $N-1$ eigenvector of the matrix
$\dot x_i +gz^{-1} U_i^- +\ell_{ij}$ and the eigenvalue $z$ is finite.
This corresponds to all other sheets of the spectral curve above the 
point $\lambda =0$. 

The expansion of the matrix $\tilde M$ yields:
$$
\tilde M_{ij}=\lambda^{-1}\dot x_i \delta_{ij}+\lambda^{-1}\dot x_i
(1-\delta_{ij}) +gz^{-1} \lambda^{-1} U_i^+ +m_{ij} +O(\lambda )
$$
$$
=\lambda^{-1}\dot x_i 
+gz^{-1} \lambda^{-1} U_i^+ +m_{ij} +O(\lambda ),
$$
where $m_{ij}$ is some matrix. On the distinguished sheet we have:
\beq\label{psi2}
\p_t \tilde c_i=
\sum_j \tilde M_{ij}\tilde c_j = z\tilde c_i +O(1), \quad z\to \infty ,
\eeq
so $c_i(t)\propto e^{tz}$ on the distinguished sheet as $z\to \infty$.
On all other sheets we have 
$\displaystyle{\sum_j \tilde M_{ij}\tilde c_j =O(1)}$ as $\lambda \to 0$.
Therefore, the $\Psi$-function has the essential singularity at the 
point $P_{\infty}$ of the form $z^{x/\eta} e^{tz}$, where $z^{-1}$ is
chosen as a local parameter around this point. On all other sheets
above $\lambda =0$ the $\Psi$-function is regular. Indeed, 
on these sheets $\displaystyle{\sum_j c_j=0}$ and so the possible poles at
$\lambda =0$ are absent.

The analysis in the neighborhood of the point $\lambda =2N\eta$
is more involved. To perform it, we need a non-trivial matrix identity
for elliptic Cauchy matrices. Let us introduce the matrix 
$G^{(n)}(\lambda )$ with matrix elements
\beq\label{psi3}
G_{ij}^{(n)}(\lambda )=\sigma (n\eta )\prod_{k\neq i}
\frac{\sigma (x_{ik}+n\eta )}{\sigma (x_{ik})}\, \phi (x_{ij}+n\eta ,
\lambda ).
\eeq
Note that $G^{(0)}(\lambda )=I$. In Appendix B we prove the
following identity:
\beq\label{psi4}
G^{(n)}(\lambda )G^{(m)}(\lambda -mN\eta )=G^{(n+m)}(\lambda -mN\eta ).
\eeq
In particular, at $m=-n$ this identity states that
\beq\label{psi5}
(G^{(n)}(\lambda ))^{-1}=G^{(-n)}(\lambda +nN\eta ).
\eeq
Using the identity (\ref{psi4}), it is not difficult to see that
\beq\label{psi6}
\Bigl (\tilde L(z,\lambda )-zI\Bigr )G^{(2)}(\lambda \! -\! 2N\eta )=D
\Bigl (L^+(z,\lambda \! -\! 2N\eta )+z^{-1}I\Bigr ),
\eeq
where $D$ is the diagonal matrix
$$
D_{ij}=\delta_{ij}\prod_{k}\frac{\sigma (x_{ik}+
\eta )}{\sigma (x_{ik}
-\eta )}
$$
and
\beq\label{psi7}
L^+_{ij}(z,\lambda )
=\dot x_i \phi (x_{ij}+\eta , \lambda )+gz
U_i^+ \phi (x_{ij}+2\eta , \lambda ).
\eeq
We see that $L^+(z,\lambda )$ has the same structure as 
$\tilde L(z, \lambda )$ and we have
\beq\label{psi1a}
L_{ij}^+(z, \lambda )=\lambda^{-1}\dot x_i +gz\lambda^{-1}
U_i^{+} +\ell^+_{ij} +O(\lambda ), \quad \lambda \to 0.
\eeq
Therefore, on the distinguished sheet we have
$$
\tilde c_i(\lambda )=\sum_jG_{ij}^{(2)}(\lambda -2N\eta )\dot x_j +
O(1), \quad \lambda \to 2N\eta ,
$$
or
\beq\label{psi8}
\tilde c_i(\lambda )=g\prod_{k\neq i} 
\frac{\sigma (x_{ik}+2\eta )}{\sigma (x_{ik})} \, 
\frac{\sum\limits_{l}\dot x_l}{\lambda -2N\eta }+O(1),
\quad \lambda \to 2N\eta .
\eeq
The action of $\tilde M$ to the vector $\tilde {\bf c}$ on all sheets
above the point $\lambda =2N\eta$ except 
the distinguishes one gives regular expressions in $z\to 0$. The action
of the singular part of $\tilde M$ as $z\to 0$ to the vector
$\tilde {\bf c}$ on the distinguished sheet yields:
$$
gz^{-1}\sum_j U_i^+ \phi (x_{ij}-\eta , \lambda )\tilde c_j =
\frac{gz^{-1}}{\sigma (-\eta )}
\sum_{j,l}\prod_{k\neq i} 
\frac{\sigma (x_{ik}+2\eta )}{\sigma (x_{ik}+\eta )}\,
G_{ij}^{(-1)}(\lambda )G^{(2)}_{jl}(\lambda -2N\eta )\dot x_l
$$
$$
=\frac{gz^{-1}}{\sigma (-\eta )}
\sum_{j,l}\prod_{k\neq i} 
\frac{\sigma (x_{ik}+2\eta )}{\sigma (x_{ik}+\eta )}\,
G_{il}^{(1)}(\lambda -2N\eta )\dot x_l
$$
$$
=-gz^{-1}\sum_{l}\prod_{k\neq i} 
\frac{\sigma (x_{ik}+2\eta )}{\sigma (x_{ik})}\, \phi (x_{il}+\eta ,
\lambda -2N\eta )\dot x_l
$$
$$
=-gz^{-1}\prod_{k\neq i} 
\frac{\sigma (x_{ik}+2\eta )}{\sigma (x_{ik})} \, 
\frac{\sum\limits_{l}\dot x_l}{\lambda -2N\eta }+O(1) =-z^{-1}\tilde c_i
+O(1),
\quad z\to 0,
$$
where in the second equality we have used the identity (\ref{psi4}).
We see that
$$
\p_t \tilde c_i =-z^{-1}\tilde c_i, \quad z\to 0
$$
on the distinguished sheet, so $c_i(t)\propto e^{-z^{-1}t}$ as $z\to 0$.
Therefore,  
the $\Psi$-function has the essential singularity at the 
point $P_{0}$ of the form $z^{x/\eta} e^{-tz^{-1}}$, where $z^{}$ is
chosen as a local parameter around this point. On all other sheets
above $\lambda =2N\eta$ the $\Psi$-function is regular.

The established analytic properties of the $\Psi$-function (\ref{a3})
allow one to conclude that it is the Baker-Akhiezer function on the 
spectral curve.

\section{Elliptic families and field generalization of the deformed
RS system}

In this section we consider more general elliptic solutions 
of the Toda hierarchy with the constraint of type B, which
are elliptic functions of a
linear combination $\displaystyle{u=\sum_k \beta_k t_k}$ 
of higher times $t_k$ of the hierarchy. In \cite{AKV02}, where
elliptic solutions to the KP hierarchy were considered, 
such solutions were called elliptic families. The dynamics 
of their poles provide field extensions of the systems of CM and RS type.
We are going to obtain the field extension of the deformed RS model
in this way. 

It is convenient to pass to the lattice space variable $n=x/\eta$ and
consider poles of the more general elliptic solutions $u_i^n$ 
(zeros of the tau-function) as functions
of the space variable $n$ and time variable $t=t_1$ 
(with $\bar t_1=-t_1$).
As it follows from the general arguments of \cite{AKV02} (see also 
\cite{ZZ21}), the general form of the tau-function is
\beq\label{f3}
\tau^n (u , t)=\rho^n (t) e^{c_1u +c_2u^2}
\prod_{j=1}^N \sigma (u - u_j^n (t)),
\eeq
where $c_1, \, c_2$ are constants and the function 
$\rho^n$ does not depend on $u$.
We assume that all $u_j^n$'s are distinct.
The linear problem (\ref{a1c}) reads
\beq\label{f4}
\p_t \Psi^{n-1}=\Psi^n +b^n \Psi^{n-1} -U^{-,n}\Psi^{n-2},
\eeq
where
\beq\label{f5}
b^n = \p_t \log \frac{\tau^n}{\tau^{n-1}}=
\sum_j \Bigl (\dot u_j^{n-1}\zeta (u-u_j^{n-1})-
\dot u_j^{n}\zeta (u-u_j^{n})\Bigr ) +\frac{\dot \rho^n}{\rho^n}-
\frac{\dot \rho^{n-1}}{\rho^{n-1}},
\eeq
\beq\label{f6}
U^{- , n}=\frac{\rho^{n- 2}\rho^{n+ 1}}{\rho^{n- 1}\rho^n}
\prod_i \frac{\sigma (u-u_i^{n- 2})
\sigma (u-u_i^{n+ 1})}{\sigma (u-u_i^{n- 1})\sigma (u-u_i^n)}.
\eeq
We assume that $U^{- , n}$ is an elliptic function of $u$, i.e.,
\beq\label{f7}
\sum_i (u_i^{n-2}+u_i^{n+1})=\sum_i (u_i^{n-1}+u_i^n).
\eeq
The pole ansatz for the $\Psi$-function is
\beq\label{f8}
\Psi^n = \sum_j c_j^n \Phi (u-u_j^n, \lambda ),
\eeq
where $\Phi (u, \lambda )$ is the Krichever function (\ref{Phi}).
Plugging (\ref{f5}), (\ref{f6}) and (\ref{f8}) into (\ref{f4}),
we obtain an equality for elliptic functions of $u$. The both sides
have poles at the points $u=u_i^{n}$ (first order poles)
and $u=u_i^{n-1}$ (second and first order poles). 
The second order poles cancel 
identically. Equating the residues at the poles, one obtains a system
of linear equations for the column vector ${\bf c}^{n} 
=(c_1^n, \ldots , c^n_N)^T$. This system can be written in the form
\beq\label{f9}
\left \{ \begin{array}{l}
{\bf c}^{n+1}=L^n {\bf c}^n +H^n {\bf c}^{n-1},
\\ \\
\dot {\bf c}^{n}=M^n {\bf c}^n -G^n {\bf c}^{n-1},
\end{array} \right.
\eeq
where $L^n$, $M^n$, $H^n$, $G^n$ are $N\! \times \! N$ matrices of the form
\beq\label{f10}
\begin{array}{l}
\phantom{a}L_{ij}^n=\dot u_i^{n+1}\Phi (u_i^{n+1}-u_j^n, \lambda ),
\\ \\
\begin{array}{l}
M_{ij}^n=(1-\delta_{ij})\dot u_i^{n}\Phi (u_i^{n}-u_j^n, \lambda )
\\ \\
\displaystyle{\phantom{aaaaaaaaaaa}
+\delta_{ij}\Bigl [ \frac{\dot \rho^{n+1}}{\rho^{n+1}}-
\frac{\dot \rho^{n}}{\rho^{n}}+\sum_{k\neq i}\dot u_k^n \zeta (u_i^n-u_k^n)
-\sum_k \dot u_k^{n+1}\zeta (u_i^n - u_k^{n+1})\Bigr ]},
\end{array}
\\ \\
\phantom{a}
\displaystyle{H_{ij}^n =\frac{\rho^{n+2}\rho^{n-1}}{\rho^{n+1}\rho^n}
\prod_k \frac{\sigma (u_i^{n+1}-u_k^{n-1})
\sigma (u_i^{n+1}-u_k^{n+2})}{\sigma (u_i^{n+1}-u_k^{n})
\sigma ^{(')}(u_i^{n+1}-u_k^{n+1})}
\, \Phi (u_i^{n+1} \! -\! u_j^{n-1}, \lambda )},
\\ \\
\phantom{a}
\displaystyle{G_{ij}^n =\frac{\rho^{n+2}\rho^{n-1}}{\rho^{n+1}\rho^n}
\prod_k \frac{\sigma (u_i^{n}-u_k^{n+2})
\sigma (u_i^{n}-u_k^{n-1})}{\sigma (u_i^{n}-u_k^{n+1})
\sigma ^{(')}(u_i^{n}-u_k^{n})}\, \Phi (u_i^n \! -\! 
u_j^{n-1}, \lambda )}.
\end{array}
\eeq
The notation $\sigma ^{(')}(u_i^{n}-u_k^{n})$ means that this 
multiplier at $k=i$ should be omitted. To avoid misunderstanding,
we should stress that the matrices $L^n$, $M^n$ differ from $L,M$ 
used in Section 2.

The system (\ref{f9}) is overdetermined. To obtain the compatibility
condition, we take the time derivative of the first equation in 
(\ref{f9}), shift $n\to n+1$ in the second one and equate the results. 
In this way, we obtain the compatibility condition:
\beq\label{f11}
\begin{array}{l}
(\dot L^n+L^nM^n-M^{n+1}L^n +G^{n+1}){\bf c}^n
\\ \\
\phantom{aaaaaaaaaaa}
+(\dot H^n +H^nM^{n-1}-M^{n+1}H^n -L^n G^n){\bf c}^{n-1}
-H^nG^{n-1}{\bf c}^{n-2}=0.
\end{array}
\eeq
Note that $L^n$, $M^n$ are the matrices of the Lax pair for the 
field analogue of the RS model (see \cite{ZZ21}). In \cite{ZZ21}),
the following matrix identity was proved:
\beq\label{f12}
\dot L^n+L^nM^n-M^{n+1}L^n =D^{n+1}L^n,
\eeq
where $D^n$ is the diagonal matrix with diagonal matrix elements
\beq\label{f13}
\begin{array}{l}
\displaystyle{
D^n_{ii}=\frac{\ddot u_i^n}{\dot u_i^n}+\sum_k \dot u_i^{n+1}
\zeta (u_i^n -u_k^{n+1})+\sum_k \dot u_i^{n-1}
\zeta (u_i^n -u_k^{n-1})}
\\ \\
\displaystyle{\phantom{aaaaaaaaaaaaaaaaaaaaaaa}-2\sum_{k\neq i}\dot u_i^{n}
\zeta (u_i^n -u_k^{n})+2\frac{\dot \rho^n}{\rho^n}-
\frac{\dot \rho^{n+1}}{\rho^{n+1}}-\frac{\dot \rho^{n-1}}{\rho^{n-1}}.}
\end{array}
\eeq
The proof is based on the identity (\ref{Phi3}) for the Krichever function.
Using this identity, we represent the first term in (\ref{f11}) in the form
$$
\Bigl (\dot L^{n-1}+L^{n-1}M^{n-1}-M^{n}L^{n-1} +G^{n}\Bigr )_{ij}
$$
$$
=\left (\tilde D^n_{ii}\dot u_i^n -W_i^{-, n}\right )
\Phi (u_i^n -u_j^{n-1}, \lambda ),
$$
where
$$
\tilde D^n_{ii}=D^n_{ii}+ (\dot u_i^n)^{-1} (W_i^{+, n}+W_i^{-, n})
$$
and
\beq\label{W}
W_i^{\pm , n}=\frac{\rho^{n\pm 2}\rho^{n\mp 1}}{\rho^{n\pm 1}\rho^n}
\prod_k \frac{\sigma (u_i^n -u_k^{n\pm 2})
\sigma (u_i^n -u_k^{n\mp 1})}{\sigma (u_i^n -u_k^{n\pm 1})
\sigma^{(')} (u_i^n -u_k^{n})}.
\eeq
Let us introduce the matrix $W^n$ with matrix elements
\beq\label{f14}
W^n_{ij}=W_i^{-, n+1}\Phi (u_i^{n+1}-u_j^n, \lambda ).
\eeq
The following matrix identities hold:
\beq\label{f15}
H^n G^{n-1} +W^n H^{n-1}=0,
\eeq
\beq\label{f16}
\dot H^n +H^n M^{n-1}-M^{n+1}H^n -W^n L^{n-1}=0.
\eeq
The proof is based on the identity (\ref{Phi3}). This identity implies that 
the matrix
element $(ij)$ of the left hand side of (\ref{f15}) 
equals $\Phi (u_i^{n+1}-
u_j^{n-2}, \lambda )$ times sum of residues of the elliptic function
$$
\Bigl (\zeta (x-u_j^{n-2})-\zeta (x-u_i^{n+1})-\zeta (u_i^{n+1}-
u_j^{n-2})+\zeta (\lambda )\Bigr )
\prod_k \frac{\sigma (x-u_k^{n-2})
\sigma (x-u_k^{n+1})}{\sigma (x-u_k^{n-1})\sigma (x-u_k^{n})}
$$
and thus is equal to zero. The calculations which are
necessary to prove (\ref{f16}) are more involved but,
again, they are based on the identity (\ref{Phi3}) (and (\ref{Phi2})).
In the calculations, one should use the fact that sum of residues
of the elliptic function
$$
\Bigl (\zeta (x-u_j^{n-1})-\zeta (x-u_i^{n+1})\Bigr )
\prod_k \frac{\sigma (x-u_k^{n+2})
\sigma (x-u_k^{n-1})}{\sigma (x-u_k^{n+1})\sigma (x-u_k^{n})}
$$
is equal to zero (note that this
function has simple as well as double poles).

Using the identities (\ref{f12}), (\ref{f15}), (\ref{f16}), we can
represent the left hand side of compatibility condition (\ref{f11})
in the form
$$
\tilde D^{n+1}L^{n}{\bf c}^n -W^n \Bigl (
{\bf c}^n -L^{n-1}{\bf c}^{n-1}-H^{n-1}{\bf c}^{n-2}\Bigr )=0.
$$
However, the expression in the brackets 
here vanishes due to the first equation
in (\ref{f9}) and we arrive at the compatibility condition in the form
$\tilde D_{ii}^n=0$ for all $i$. This gives the equations of motion
for the lattice ``fields'' $u_i^n$:
\beq\label{f17}
\begin{array}{c}
\displaystyle{
\ddot u_i^n +\sum_j \Bigl [ \dot u_i^n \dot 
u_j^{n+1}\zeta (u_i^n -u_j^{n+1})
+\dot u_i^n \dot u_j^{n-1}\zeta (u_i^n -u_j^{n-1})\Bigr ]
-2\sum_{j\neq i}\dot u_i^n \dot u_j^{n}\zeta (u_i^n -u_j^{n})}
\\ \\
\displaystyle{
-\, \dot u_i^n \p_t \log \frac{\rho^{n+1}\rho^{n-1}}{(\rho^n)^2}\,
+W_i^{+, n}+W_i^{-, n}=0,}
\end{array}
\eeq
where $W_i^{\pm , n}$ are given by (\ref{W}). Summing them over all 
$i$, we obtain the equation
\beq\label{f18}
\begin{array}{l}
\displaystyle{
\p_t \log \frac{\rho^{n+1}\rho^{n-1}}{(\rho^n)^2}\,
\sum_i \dot u_i^n}
\\ \\
\phantom{aaaaaaa}
\displaystyle{ = \, \sum_{i,j}\Bigl [ \dot u_i^n \dot 
u_j^{n+1}\zeta (u_i^n -u_j^{n+1})
+\dot u_i^n \dot u_j^{n-1}\zeta (u_i^n -u_j^{n-1})\Bigr ]
+\sum_i (W_i^{+, n}+W_i^{-, n}).}
\end{array}
\eeq
Equations (\ref{f17}) together with (\ref{f18}) form a system of
$N+1$ differential equations for the $N+1$ fields $u_j^n=u_j^n(t)$ ($j=1, 
\ldots , N$), $\rho^n =\rho^n(t)$. These equations provide the field 
extension of the deformed RS system. They were obtained in \cite{Z23}
by a different method. Our new method allows us to provide a commutation
representation for them in the form of a field extension of the
Manakov triple:
\beq\label{f19}
\dot L^n +L^n M^n -M^{n+1}L^n +R^n L^n =0,
\eeq
where $R^n$ is the diagonal matrix with matrix elements
$$
R^n_{ii}=(\dot u_i^{n+1})^{-1}(W_i^{+, \, n+1}+W_i^{-, \, n+1}).
$$
If we set $u_i^n=x_i-n\eta$, equations (\ref{f17}), (\ref{f18}) reduce
to the equations of motion of the deformed RS system for the $x_i$'s.

\section{Open problems}

In conclusion,
let us list some open problems and ideas for further research 
related to the 
deformed RS system. 

\begin{itemize} 
\item[a)]
The most important unsolved problem is to find the Hamiltonian structure
of the deformed RS model, if any.
With the commutation representation 
at hand, one may hope to apply the general method developed by Krichever
to approach this problem. A related question is 
quantization of the deformed
RS system.

\item[b)]
In \cite{Z23} an integrable time discretization of the deformed RS system
was obtained. It is desirable to find a commutation representation 
of the fully discrete equations of motion. Presumably, the pole ansatz
for elliptic solutions of the fully discrete BKP equation should be used.

\item[c)]
It is known that the rational and trigonometric CM and RS models are
connected by various duality relations. It would be interesting to
find similar duality relations for the deformed RS models and their
$\eta \to 0$ degenerations. 

\item[d)]
A natural generalization of our result would be the extension
of the correspondence with dynamics of poles of elliptic solutions
to the whole infinite hierarchies, as it was done for the Toda/RS 
hierarchies in \cite{PZ21}.

\item[e)]
It is known that the field generalization of the RS model
is gauge equivalent to a classical spin model on the lattice \cite{ZZ21}.
It is natural to ask whether any models of the spin chain type are
equivalent in this sense to the field extension of the deformed RS model
obtained in this paper. 

\end{itemize}

\section*{Appendix A: The Weierstrass and Krichever functions}
\addcontentsline{toc}{section}{Appendix A: The Weierstrass 
and Krichever functions}
\def\theequation{A\arabic{equation}}
\def\theHequation{\theequation}
\setcounter{equation}{0}

In this appendix we present the definition and main properties of the 
special functions used in the main text.

Let $\omega$, $\omega '$ be complex numbers such that 
${\rm Im} (\omega '/ \omega )>0$.
The Weierstrass $\sigma$-function 
with quasi-periods $2\omega$, $2\omega '$ 
is defined by the following infinite product over the lattice
$2\omega m+2\omega ' m'$, $m,m'\in \ZZ$:
\beq\label{A1}
\sigma (x)=\sigma (x |\, \omega , \omega ')=
x\prod_{s\neq 0}\Bigl (1-\frac{x}{s}\Bigr )\, 
e^{\frac{x}{s}+\frac{x^2}{2s^2}},
\quad s=2\omega m+2\omega ' m' \quad m, m'\in \ZZ .
\eeq 
It is an odd quasiperiodic function with two linearly independent
quasi-periods in the complex plane. 
The expansion around $x=0$ is
\beq\label{A1a}
\sigma (x)=x+O(x^5), \quad x\to 0.
\eeq
The monodromy properties of the $\sigma$-function 
under shifts by the quasi-periods
are as follows:
\beq\label{A4}
\begin{array}{l}
\sigma (x+2\omega )=-e^{2\zeta (\omega )(x+\omega )}\sigma (x),
\\ \\
\sigma (x+2\omega ' )=-e^{2\zeta (\omega ')(x+\omega ' )}\sigma (x).
\end{array}
\eeq
Here $\zeta (x)$ is the
Weierstrass $\zeta$-function defined as
\beq\label{A4a}
\zeta (x)=\frac{\sigma '(x)}{\sigma (x)}.
\eeq
As $x\to 0$,
\beq\label{A2a}
\zeta (x)=\frac{1}{x} +O(x^3), \quad x\to 0.
\eeq

The Weierstrass $\wp$-function is defined as
$\wp (x)=-\zeta '(x)$. 
It is an even double-periodic function with periods $2\omega , 2\omega '$
and with second order poles at the points 
of the lattice $s=2\omega m+2\omega ' m'$ with integer $m, m'$.
As $x\to 0$, $\wp (x)=x^{-2}+O(x^2)$.

In the main text we use the Krichever function
\beq\label{Phi}
\Phi (x, \lambda )=
\frac{\sigma (x+\lambda )}{\sigma (\lambda )\sigma (x)}\,
e^{-\zeta (\lambda )x}
\eeq
which has a simple pole
at $x=0$. 
The expansion of $\Phi (x, \lambda )$ as $x\to 0$ is
$$
\Phi (x, \lambda )=\frac{1}{x}+\alpha_1 x +\alpha_2 x^2 +\ldots , \qquad 
x\to 0,
$$
where $\alpha_1=-\frac{1}{2}\, \wp (\lambda )$, 
$\alpha_2=-\frac{1}{6}\, \wp '(\lambda )$. 
The quasiperiodicity properties of the Krichever function are:
\beq\label{quasi}
\begin{array}{l}
\Phi (x+2\omega , \lambda )=e^{2(\zeta (\omega )\lambda - 
\zeta (\lambda )\omega )}
\Phi (x, \lambda ),
\\ \\
\Phi (x+2\omega ' , \lambda )=e^{2(\zeta (\omega ' )\lambda - \zeta (\lambda )\omega ' )}
\Phi (x, \lambda ).
\end{array}
\eeq
As a function of $\lambda$, $\Phi (x, \lambda )$ is a double-periodic
function with periods $2\omega$, $2\omega '$.

In the main text we use the following identities for the 
Krichever function:
\beq\label{Phi2}
\p_x \Phi (x, \lambda )=\Phi (x, \lambda )\Bigl (\zeta (x+\lambda )-
\zeta (x)-\zeta (\lambda )\Bigr ),
\eeq
\beq\label{Phi3}
\Phi (x, \lambda )\Phi (y, \lambda )=\Phi (x+y, \lambda )
\Bigl (\zeta (x)+\zeta (y)-
\zeta (x+y+\lambda )+\zeta (\lambda )\Bigr ).
\eeq
The first of them directly follows from the definition.
To prove the second one, consider the function
$\Phi (x, \lambda )\Phi (y, \lambda )/\Phi (x+y, \lambda )$ as a function
of $x$. It is an elliptic function of $x$ whose poles and zeros 
coincide with those of the function 
$\zeta (x)+\zeta (y)-
\zeta (x+y+\lambda )+\zeta (\lambda )$. Therefore, their ratio 
is a constant
which can be found
by putting $x$ to some special value. 

\section*{Appendix B: Proof of the matrix identity}
\addcontentsline{toc}{section}{Appendix B: Proof of the matrix identity}
\def\theequation{B\arabic{equation}}
\def\theHequation{\theequation}
\setcounter{equation}{0}

Let $G^{(n)}(\lambda )$ be the $N\! \times \! N$ 
matrix with matrix elements
\beq\label{B1}
G_{ij}^{(n)}(\lambda )=\sigma (n\eta )\prod_{k\neq i}
\frac{\sigma (x_{ik}+n\eta )}{\sigma (x_{ik})}\, \phi (x_{ij}+n\eta ,
\lambda ),
\eeq
where
$$
\phi (x, \lambda )=\frac{\sigma (x+\lambda )}{\sigma (\lambda )
\sigma (x)}.
$$
The purpose of this appendix is to prove the
following matrix identity:
\beq\label{B2}
G^{(n)}(\lambda )G^{(m)}(\lambda -mN\eta )=G^{(n+m)}(\lambda -mN\eta ).
\eeq

The proof is by induction in $N$. A simple calculation shows that
(\ref{B2}) holds at $N=1$. The induction assumption is 
that it is true for all matrices
up to size $N-1$. Let us show that it holds for $N\! \times \! N$ matrices.
We will prove that
$$
S_{ik}:=\frac{(G^{(n)}(\lambda )
G^{(m)}(\lambda -mN\eta ))_{ik}}{(G^{(n+m)}(\lambda -mN\eta ))_{ik}}
$$
is a double-periodic function in $\lambda$ and $x_i$ without poles. 
Whence it is a constant which can be found by putting $x_i$ to some
particular value. Explicitly, we have:
\beq\label{B3}
\begin{array}{c}
\displaystyle{
S_{ik}=\frac{\sigma (n\eta )\sigma (m\eta )}{\sigma ((n+m)\eta )}
\left (\prod_{l\neq i} \frac{\sigma (x_{il}+n\eta )}{\sigma (x_{il}
+(n+m)\eta )}\right )}
\\ \\
\displaystyle{
\times \, \sum_j \frac{\sigma (x_{ij}\! +\! n\eta \! +\! \lambda )
\sigma (x_{jk}\! -\! m(N\! -\! 1)\eta \! +\! \lambda )
\sigma (x_{ik}\! +\! (m\! +\! n)\eta )}{\sigma (x_{ik}\! +\! 
(m\! +\! n\! -\! mN)\eta \! +\! \lambda )\sigma (\lambda )
\sigma (x_{ij}\! +\! n\eta )\sigma (x_{jk}\! +\! m\eta )}
\left (\prod_{s\neq j}\frac{\sigma (x_{js}\! +\! m\eta )}{\sigma (x_{js})}
\right )}.
\end{array}
\eeq

First, let us consider this function as a function of $\lambda$. 
Obviously, it is double-periodic in $\lambda$ with possible simple poles
at $\lambda =0$ and $\lambda =-x_{ik}-(m+n-mN)\eta$ if 
$x_{ik}+(m+n-mN)\eta \neq 0$. The residue is proportional to
$$
r=\sigma (m\eta )\sum_j \frac{\sigma (x_{jk}-m(N-1)\eta )}{\sigma
(x_{jk}+m\eta )}\left (\prod_{s\neq j}
\frac{\sigma (x_{js}\! +\! m\eta )}{\sigma (x_{js})}
\right )=0
$$
since it is the sum of residues of the elliptic function
$$
r(x)=\frac{\sigma (x-x_k -m(N-1)\eta )}{\sigma (x-x_k +m\eta )}
\left (\prod_s \frac{\sigma (x-x_s +m\eta )}{\sigma (x-x_s)}\right ).
$$
Hence $S_{ik}$ is regular as a function of $\lambda$. 
At $x_{ik}+(m+n-mN)\eta =0$ we have a double pole with a coefficient
proportional to $r$, hence this pole is actually 
absent. Therefore, we conclude
that $S_{ik}$ does not depend on $\lambda$. 

Next, consider $S_{ik}$ as a function of $x_i$. From (\ref{B3}) it is clear
that it is an elliptic function of $x_i$. Let us consider possible
poles of this function:
\begin{itemize}
\item[a)] A possible simple pole at $x_i=x_k-(m+n-mN)\eta -\lambda$ which
depends on $\lambda$. Since $S_{ik}$ does not depend on $\lambda$, this 
pole is actually absent. 
\item[b)] Possible simple poles at $x_i=x_l$. The residue is proportional to
$$
\frac{\sigma (m\eta )\sigma (n\eta +\lambda )\sigma (x_{lk}-m(N-1)\eta 
+\lambda )}{\sigma (n\eta )\, \sigma (x_{lk}+m\eta )}
\prod_{s\neq l,i}\frac{\sigma (x_{ls}+m\eta )}{\sigma (x_{ls})}
$$
$$
-\, \frac{\sigma (m\eta )\sigma (n\eta +\lambda )\sigma (x_{ik}-m(N-1)\eta 
+\lambda )}{\sigma (n\eta )\, \sigma (x_{ik}+m\eta )}
\prod_{s\neq l,i}\frac{\sigma (x_{is}+m\eta )}{\sigma (x_{is})}=0.
$$
The first term comes from the sum in 
(\ref{B3}) at $j=l$, the second one from
$j=i$. 
\item[c)] Possible simple poles at $x_i=x_l -(m+n)\eta$ with $l\neq k$.
The residue is proportional to
$$
r_l\! =\! \sum_{j\neq i} \! \frac{\sigma (x_{lj}\! -\! m\eta \! +\! \lambda )
\sigma (x_{jk}\! -\! m(N\! -\! 1)\eta \! +\! \lambda )}{\sigma
(x_{lj}\! -\! m\eta )\sigma (x_{jk}\! +\! m\eta )}\! \left (
\prod_{s\neq j,i} \frac{\sigma (x_{js}\! +\! m\eta )}{\sigma (x_{js})}
\right ) \! \frac{\sigma (x_{jl}\! +\! (2m\! +\! n)\eta )}{\sigma
(x_{jl}\! +\! (m\! +\! n)\eta )}
$$
$$
+\, \frac{\sigma (n\eta \! +\! \lambda )
\sigma (x_{lk}\! -\! (n+mN)\eta \! +\! \lambda )}{\sigma (n\eta )\, \sigma (x_{lk}\! -\! n\eta )}\left (\prod_{s\neq i}
\frac{\sigma (x_{ls}\! -\! n\eta )}{\sigma
(x_{ls}\! -\! (n+m)\eta )}\right ).
$$
As soon as it does not depend on $\lambda$ we can put $\lambda =-n\eta$,
then we have
$$
r_l = \sum_{j\neq i}  \frac{\sigma (x_{jl}\! +\! (2m\! +\! n)
\eta )\sigma (x_{kj}\! -\! (n\! +\! m(N\! -\! 1))\eta )}{\sigma
(x_{jl} + m\eta )\sigma (x_{kj} - m\eta )}
\prod_{s\neq j,i} \frac{\sigma (x_{js}\! +\! m\eta )}{\sigma (x_{js})}.
$$
Rearranging, we write this as
$$
r_l=\sigma ((n+mN)\eta )\sigma ((n+m)\eta )\prod_{s\neq j,i}
\frac{\sigma (x_{js}+m\eta )}{\sigma (x_{js})}
$$
$$
\times \sum_{j\neq i} \frac{\sigma (x_{kj}\! -\! m\eta \! +\! (n\! +\! 
mN)\eta )}{\sigma ((n+mN)\eta )\sigma (x_{kj}-m\eta )}\,
\frac{\sigma (x_{jl}\! +\! m\eta \! +\! (n\! +\! mN)\eta \! -\! 
m(N\! -\! 1)\eta )}{\sigma ((n+m)\eta )\sigma (x_{jl}+m\eta )}.
$$
This is proportional to the matrix element of the 
product $$G^{(-m)}(\lambda )G^{(m)}(\lambda -m(N-1)\eta )$$ for 
$(N-1)\times (N-1)$ matrices at $\lambda =(n+mN)\eta$ which is 
the unity matrix by the induction assumption. Therefore, we have:
$$
r_l=-\delta_{kl} \frac{\sigma ((n+mN)\eta )\sigma ((n+m)\eta )}{\sigma ^2
(m\eta )}\prod_{s\neq k,i} \frac{\sigma (x_{ks})}{\sigma (x_{ks}-m\eta )}=0
$$
since $l\neq k$.
\end{itemize}
We conclude that $S_{ik}$ is regular as a function of $x_i$ and thus
it does not depend on $x_i$. To find it, we put $x_i=x_l-n\eta$ for some
$l\neq i$. The evaluation of $S_{ik}$ at this point yields $S_{ik}=1$,
so the identity (\ref{B2}) is proved. 

\section*{Acknowledgments}

\addcontentsline{toc}{section}{Acknowledgments}

The research of A.Z. has been supported 
in part within the framework of the
HSE University Basic Research Program.

\end{document}